\documentclass[twoside,12pt]{article}
\usepackage{CJK}
\usepackage{indentfirst}
\usepackage{bm}
\usepackage{verbatim}
\usepackage{graphicx}
\usepackage{multicol}

\begin{document}
\begin{CJK*}{GBK}{song}
\newcommand{\song}{\CJKfamily{song}}
\newcommand{\hei}{\CJKfamily{hei}}
\newcommand{\fs}{\CJKfamily{fs}}
\newcommand{\kai}{\CJKfamily{kai}}
\def\thefootnote{\fnsymbol{footnote}}
\begin{center}
\Large\hei  The quantum entropic uncertainty relation and entanglement witness in the two-atom system coupling with the non-Markovian environments   
\end{center}

\footnotetext{\hspace*{-.45cm}\footnotesize $^*$zhmzc1997@126.com; $^\dag$Corresponding author:mffang@hunnu.edu.cn, tel:13973181728}

\begin{center}
\rm Hong-Mei Zou$^{*}$, \ Mao-Fa Fang$^{\dag}$, Bai-Yuan Yang, You-Neng Guo, Wei He, Shi-Yang Zhang
\end{center}

\begin{center}
\begin{footnotesize} \rm
Key Laboratory of Low-dimensional Quantum Structures and Quantum Control of Ministry of Education, College of Physics and Information Science, Hunan Normal University, Changsha, 410081, China\  \\   

\end{footnotesize}
\end{center}

\begin{center}
\footnotesize (Received 23 3 2014; revised manuscript received X XX
XXXX)
\end{center}

\vspace*{2mm}

\begin{center}
\begin{minipage}{15.5cm}
\parindent 20pt\footnotesize

The quantum entropic uncertainty relation and entanglement witness in the two-atom system coupling with the non-Markovian environments are studied by the time-convolutionless master-equation approach. The influence of non-Markovian effect and detuning on the lower bound of the quantum entropic uncertainty relation and entanglement witness is discussed in detail. The results show that, only if the two non-Markovian reservoirs are identical, increasing detuning and non-Markovian effect can reduce the lower bound of the entropic uncertainty relation, lengthen the time region during which the entanglement can be witnessed, and effectively protect the entanglement region witnessed by the lower bound of the entropic uncertainty relation. The results can be applied in quantum measurement, quantum cryptography task and  quantum information processing.

\end{minipage}
\end{center}

\begin{center}
\begin{minipage}{15.5cm}
\begin{minipage}[t]{2.3cm}{\hei Keywords: }\end{minipage}
\begin{minipage}[t]{13.1cm} entropic uncertainty relation, quantum memory, entanglement witness, non-Markovian environment
\end{minipage}\par\vglue8pt
{\bf PACS: }03.67.Mn, 03.67.Dd, 03.65.Yz, 89.70.cf
\end{minipage}
\end{center}

\section{\fs Introduction}  
The entropy uncertainty relation and its application have received plenty of attention in quantum optics and quantum information processing, currently. The uncertainty principle is a core of quantum mechanics and remarkably illustrates the difference between classical and quantum mechanics. There are two different formulae for the uncertainty principle: one is the Heisenberg uncertainty relation\cite{Heisenberg}, which measures the quantum fluctuations of two observables by standard deviations, i.e. $\triangle R\cdot\triangle Q \geq \frac{1}{2} |\langle [R,Q]\rangle|$, for two incompatible observables $R$ and $Q$. Another is the entropic uncertainty relation\cite{Bialynicki1,Bialynicki2,Vaccaro} which quantifies the quantum fluctuations of two observables in terms of entropy. Suppose $p(x)$ is a probability distributions of the measurement outcome $x$ for a random variable $X$, the Shannon entropy $H(X)=-\sum_{k}p_{k}(x)log_{2}p_{k}(x)$ indicates the uncertainty of $X$. Thus the entropic uncertainty relation\cite{Hirschman,Deutsch,Kraus,Maassen,Sanchez} is expressed as $H(Q)+H(R)\geq log_{2}\frac{1}{c}$ for two incompatible observables $R$ and $Q$, where  $\frac{1}{c}$ is defined as the complementarity of the two observables. For non-degenerate observables, $c:=max_{j,k}|\langle\psi_{j}|\phi_{k}\rangle|^{2}$ if the eigenvectors of $Q$ and $R$ are respectively $|\psi_{j}\rangle$ and $|\phi_{k}\rangle$. Recently, some important progress has been acquired about the entropic uncertainty relation and its application. Several uncertainty relations have been proposed, one of them is the uncertainty principle in the presence of quantum memory which is put forward by Renes group\cite{Renes,Berta}. This uncertainty principle has important significance that the known quantum information stored in the quantum memory can reduce or eliminate the uncertainty about measurement outcomes of another particle which is entangled with the quantum memory, and is confirmed in recent experiments\cite{Prevedel,LiC}. The entropic uncertainty relations have been widely used in quantum entanglement witness\cite{Berta}, security analysis of quantum cryptographic protocols\cite{Tomamichel}, locking of classical correlation in quantum state\cite{DiVincenzo}, quantum phase transitions\cite{Nataf} and quantum information processing\cite{Hu1}.

For real quantum systems, which unavoidably interact with their environments\cite{Breuer1,ZouHM}, how do environments affect the entropic uncertainty relation in the presence of quantum memory? Recently, more and more attention has been paid to this topic, such as, M. Feng \emph{et al.}\cite{Xu1} explored the quantum-memory assisted entropic uncertainty relation under noises and found that the unital noises only increase the uncertainty while the amplitude-damping nonunital noises may reduce the uncertainty in the long-time limit. H. Fan \emph{et al.}\cite{Hu1} studied the relations between the quantum-memory-assisted entropic uncertainty principle, quantum teleportation and entanglement witness. A. K. Pati \emph{et al.}\cite{Pati} researched the influence of quantum discord and classical correlation on the entropic uncertainty relation in the presence of quantum memory and showed that the quantum discord and the classical correlations can tighten the lower bound of Berta \emph{et al.}\cite{Berta}. In this paper, we investigate the quantum entropic uncertainty relation and entanglement witness in the two-atom system coupling with the non-Markovian environments by the time-convolutionless master-equation approach. We propose a method to reduce the lower bound of the entropic uncertainty relation and enhance entanglement witness in non-Markovian environments.

The paper is organized as follows. In Section 2, we present the time-convolutionless master-equation of the two atoms in independent reservoirs. In Section 3,  we introduce the entropic uncertainty relation in the presence of quantum memory. Then we discuss in Section 4 the influence of non-Markovian effect and detuning on the lower bound of the entropic uncertainty relation and entanglement witness. Finally, a brief summary is given in Section 5.

\section{\fs The time-convolutionless master-equation of the two-atom system in independent reservoirs}

Suppose that two two-level atoms off-resonantly couple to two non-Markovian reservoirs with zero temperature\cite{ZouHM,Ferraro}. The total Hamiltonian that describes such a system can be written as($\hbar=1$)
\begin{eqnarray}\label{EB01}
H&=&H_{0}+\alpha H_{I},
\end{eqnarray}
where
\begin{eqnarray}\label{EB02}
H_{0}&=&\omega_{0}\sum_{j=A}^{B}S_{j}^{z}+\sum_{n}\omega_{n,A}b_{n,A}^{\dag}b_{n,A}+\sum_{m}\omega_{m,B}b_{m,B}^{\dag}b_{m,B}
\end{eqnarray}
is the free Hamiltonian of the combined system. $\omega_{0}$ is the transition frequency of the two atoms, $S_{j}^{z}$ is the inversion operators describing the atom $j$($j=A$ or $B$), $b_{n,A}^{\dag}(b_{m,B}^{\dag})$ and $b_{n,A}(b_{m,B})$ are the creation and annihilation operators of the bosonic bath with the frequency $\omega_{n,A}(\omega_{m,B})$. The parameter $\alpha$ is a dimensionless expansion parameter. The interaction Hamiltonian $H_{I}$ is given by
\begin{eqnarray}\label{EB03}
H_{I}&=&\sum_{n}g_{n,A}b_{n,A}S_{A}^{+}+\sum_{m}g_{m,B}b_{m,B}S_{B}^{+}+h.c. ,
\end{eqnarray}
where $g_{n,A}(g_{m,B})$ is the coupling constant between the atom and its corresponding reservoir, $S_{j}^{+}$ and
$S_{j}^{-}$ are the upward and downward operators of the atom, respectively. In the interaction picture, the Hamiltonian $\alpha H_{I}$ reads as
\begin{eqnarray}\label{EB04}
\alpha H_{I}(t)&=&\sum_{j=A}^{B}(S_{j}^{+}\sum_{n}g_{n,j}b_{n,j}e^{i(\omega_{0}-\omega_{n,j})t}
+S_{j}^{-}\sum_{n}g_{n,j}^{\ast}b_{n,j}^{\dagger}e^{-i(\omega_{0}-\omega_{n,j})t}).
\end{eqnarray}

In the second order approximation, the time-convolutionless(TCL) master equation\cite{Breuer1}, described by the density operator $\rho_{AB}(t)$, has the following form
\begin{eqnarray}\label{EB05}
\frac{d}{dt}\rho_{AB}{(t)}&=&-\alpha^{2}\int_{0}^{t}d\tau Tr_{E}([H_{I}{(t)},[H_{I}{(\tau)},\rho_{AB}{(t)}\otimes\rho_{E}]])
\end{eqnarray}
with the environment state $\rho_{E}$.

Here we have assumed that $\rho{(t)}=\rho_{AB}{(t)}\otimes\rho_{E}$ and
$Tr_{E}([H_{I}{(t)},\rho_{AB}{(0)}\otimes\rho_{E}])=0$,
and Eq.~(\ref{EB05}) may be written as
\begin{eqnarray}\label{EB06}
\frac{d}{dt}\rho_{AB}{(t)}&=&\mathcal{L}^{(A)}\rho_{AB}{(t)}+\mathcal{L}^{(B)}\rho_{AB}{(t)},
\end{eqnarray}
where $\mathcal{L}^{(j)}$($j=A$ or $B$) is the Liouville superoperator\cite{Sinayskiy} associated to the Hamiltonian of Eq.~(\ref{EB04}) and it is defined by
\begin{eqnarray}\label{EB07}
\mathcal{L}^{(j)}\rho_{AB}{(t)}&=&f_{j}(t)[S_{j}^{-}\rho_{AB}{(t)},S_{j}^{+}]+f_{j}^\ast(t)[S_{j}^{-},\rho_{AB}{(t)}S_{j}^{+}]\nonumber\\
&&+k_{j}^\ast(t)[S_{j}^{+}\rho_{AB}{(t)},S_{j}^{-}]+k_{j}(t)[S_{j}^{+},\rho_{AB}{(t)}S_{j}^{-}].
\end{eqnarray}
The correlation functions $k_{j}(t)$ and $f_{j}(t)$ are given by
\begin{eqnarray}\label{EB08}
k_{j}(t)&=&i\sum_{n}|g_{n,j}|^{2}\langle b_{n,j}^{\dagger}b_{n,j}\rangle_{E_{j}}\frac{1-e^{i(\omega_{0}-\omega_{n,j})t}}{\omega_{0}-\omega_{n,j}}
\end{eqnarray}
and
\begin{eqnarray}\label{EB09}
f_{j}(t)&=&i\sum_{n}|g_{n,j}|^{2}\langle b_{n,j}b_{n,j}^{\dagger}\rangle_{E_{j}}\frac{1-e^{i(\omega_{0}-\omega_{n,j})t}}{\omega_{0}-\omega_{n,j}},
\end{eqnarray}
where $\langle b_{n,j}^{\dagger}b_{n,j}\rangle_{E_{j}}=Tr_{E_{j}}(( b_{n,j}^{\dagger}b_{n,j})\rho_{E_{j}})$ and $\langle b_{n,j}b_{n,j}^{\dagger}\rangle_{E_{j}}=Tr_{E_{j}}(( b_{n,j}b_{n,j}^{\dagger})\rho_{E_{j}})$.

Presuming that the two reservoirs are initially prepared in the thermal state with zero temperature, the correlation functions reduce to
\begin{eqnarray}\label{EB10}
k_{j}(t)=0,f_{j}(t)&=&i\sum_{n}|g_{n,j}|^{2}\frac{1-e^{i(\omega_{0}-\omega_{n,j})t}}{\omega_{0}-\omega_{n,j}}.
\end{eqnarray}
For a sufficiently large environment, we can replace the sum over the discrete coupling constants with an integral over a continuous distribution of frequencies of the environmental modes, i.e. $\sum_{n}|g_{n,j}|^{2}\rightarrow\int_{0}^{\infty}d\omega_{j} J(\omega_{j})$.

We consider the $j$-th($j=A,B$) reservoir with the Lorentzian spectral density\cite{Breuer1,Breuer2}
\begin{equation}\label{EB11}
J(\omega_{j})=\frac{1}{2\pi}\frac{\gamma_{0}\lambda_{j}^{2}}{(\omega_{0}-\omega_{j}-\delta_{j})^{2}+\lambda_{j}^{2}},
\end{equation}
where $\delta_{j}$ is the detuning between $\omega_{0}$ and the center frequency $\omega_{j}$ of the j-th reservoir. And the parameter $\lambda_{j}$ defines the spectral width of the coupling, which is connected to the reservoir correlation time $\tau_{R_{j}}$ by $\tau_{R_{j}}\approx\lambda_{j}^{-1}$. On the other hand, the parameter $\gamma_{0}$ can be shown to be related to the decay of the excited state of the atom in the Markovian limit of flat spectrum. The relaxation time scale $\tau_{S}$ over which the state of the system changes is then related to $\gamma_{0}$ by $\tau_{S}\approx\gamma_{0}^{-1}$. Utilizing Eq.~(\ref{EB10}) and Eq.~(\ref{EB11}), we can obtain the correlation functions
\begin{eqnarray}\label{EB12}
k_{j}(t)=0,f_{j}(t)&=&\frac{\gamma_{0}\lambda_{j}}{2(\lambda_{j}-i\delta_{j})}[1-e^{(i\delta_{j}-\lambda_{j})t}].
\end{eqnarray}

In the subsequent analysis of dynamical evolution of the system, typically a weak and a strong coupling regimes can be distinguished. For a weak regime we mean the case $\lambda_{j}>2\gamma_{0}$, that is, $\tau_{S}>2\tau_{R_{j}}$. In this regime the relaxation time is greater than the reservoir correlation time and the behavior of dynamical evolution of the system is essentially a Markovian exponential decay controlled by $\gamma_{0}$. In the strong coupling regime, that is, for $\lambda_{j}<2\gamma_{0}$, or $\tau_{S}<2\tau_{R_{j}}$, the reservoir correlation time is greater than or of the same order as the relaxation time and non-Markovian effects become relevant. The dynamical evolution of the system will oscillate up and down due to the memory and feedback of non-Markovian environments. For this reason we are interested in this regime and we shall mainly limit our considerations to this case\cite{Breuer1,dalton,Bellomo1,Bellomo2,ChenJJ}.

\section{\fs The entropic uncertainty relation in the presence of quantum memory}
We consider the entropic uncertainty game model illustrated in Ref.\cite{Berta}. Before the game commences, Alice and Bob agree on the two measurements, the atomic polarization components $S_{x}$ and $S_{y}$. The game proceeds as follows: Bob sends an atom $A$, initially entangled with another atom $B$ (quantum memory), to Alice. Then, Alice measures $S_{x}$ and $S_{y}$, and announces her measurement choice to Bob. The entropic uncertainty relation about Alice's measurement outcomes in the presence of quantum memory is written as
\begin{eqnarray}\label{EB13}
H(S_{x}|B)+H(S_{y}|B)&\geq&log_{2}\frac{1}{c}+H(A|B),
\end{eqnarray}
where $H(A|B)$=$H(\rho_{AB})-H(\rho_{B})$, $H(S_{x}|B)$=$H(\rho_{S_{x}B})-H(\rho_{B})$ and $H(S_{y}|B)$=$H(\rho_{S_{y}B})-H(\rho_{B})$ are respectively the conditional von Neumann entropies of the states $\rho_{AB}$, $\rho_{S_{x}B}$ and $\rho_{S_{y}B}$, where $H(\rho)=-tr(\rho log_{2}\rho)=-\sum_{j}\lambda_{j}log_{2}\lambda_{j}$ is the von Neumann entropy of the state $\rho$\cite{Nielsen}, $\rho_{S_{x}B}$ and $\rho_{S_{y}B}$ are the post-measurement states after $S_{x}$ and $S_{y}$ are performed on the atom A, $\lambda_{j}$ are the eigenvalues of the state $\rho$. The significance of this entropic uncertainty relation is that the known quantum information about $A$, stored in the quantum memory $B$, can reduce or eliminate the entropic uncertainty about Alice's measurement outcomes. Thus the bigger the entanglement between $A$ and $B$ is, the smaller the lower bound of the entropic uncertainty relation is. The term $\frac{1}{c}$ quantifies the complementarity of the two observables. If $|\psi_{j}\rangle$ and $|\phi_{k}\rangle$ are respectively the eigenvectors of $S_{x}$ and $S_{y}$, $c:=max_{j,k}|\langle\psi_{j}|\phi_{k}\rangle|^{2}=\frac{1}{2}$. The results in Refs.\cite{Cerf,Vollbrecht,Devetak} show that a negative conditional entropy is a signature of entanglement, i.e. $\rho_{AB}$ is entangled when $H(A|B)<0$. Hence the entanglement can be witnessed by the right-hand side of the inequality in Eq.~(\ref{EB13}). That is, $A$ is entangled with $B$ if $log_{2}\frac{1}{c}+H(A|B)<1$.

Let that the initial state of the two atoms is
\begin{eqnarray}\label{EB17}
|\Psi(0)\rangle_{AB}&=&\frac{1}{\sqrt{2}}(|00\rangle+|11\rangle)_{AB}.
\end{eqnarray}
Solving Eq.~(\ref{EB06}), the density matrix of the two-atom system at time $t$ has the following form
\begin{eqnarray}\label{EB14}
\rho_{AB}(t)&=&\left(
            \begin{array}{cccc}
              \rho_{11} & \rho_{12} & \rho_{13} & \rho_{14} \\
              \rho_{21} & \rho_{22} & \rho_{23} & \rho_{24} \\
              \rho_{31} & \rho_{32} & \rho_{33} & \rho_{34} \\
              \rho_{41} & \rho_{42} & \rho_{43} & \rho_{44} \\
            \end{array}
          \right).
\end{eqnarray}

After Alice measures $S_{x}$ or $S_{y}$, the system state is
\begin{eqnarray}\label{EB15}
\rho_{S_{x}B}&=&\sum_{j}(|\psi_{j}\rangle\langle\psi_{j}|\otimes I_{B})\rho_{AB}(|\psi_{j}\rangle\langle\psi_{j}|\otimes I_{B})
\end{eqnarray}
or
\begin{eqnarray}\label{EB16}
\rho_{S_{y}B}&=&\sum_{k}(|\phi_{k}\rangle\langle\phi_{k}|\otimes I_{B})\rho_{AB}(|\phi_{k}\rangle\langle\phi_{k}|\otimes I_{B}).
\end{eqnarray}

In the following, we analyze numerically the lower bound of the entropic uncertainty relation and witness entanglement according to this lower bound.

\section{\fs Results and Discussions}
Based on the formulae introduced in section 3, we can obtain numerically the lower bound of the entropic uncertainty relation and witness entanglement according to this lower bound. We employ Wootter's concurrence\cite{Wootter} quantifying the entanglement between the two atoms, which is defined as
\begin{equation}\label{EB119}
C_{AB}=max(0,\sqrt{\lambda_{1}}-\sqrt{\lambda_{2}}-\sqrt{\lambda_{3}}-\sqrt{\lambda_{4}})
\end{equation}
where $\lambda_{i}$ are the eigenvalues, organized in a descending order, of the matrix $\tilde{\rho}
=\rho_{AB}(\sigma_{y}\otimes\sigma_{y})\rho_{AB}^{\ast}(\sigma_{y}\otimes\sigma_{y})$.

For simplicity, we introduce three abbreviations. The minimum uncertainty($MU$) represents the lower bound of the entropic uncertainty relation in Eq.~(\ref{EB13}), i.e. $MU=log_{2}\frac{1}{c}+H(A|B)$. The time of entanglement witnessed($T_{EW}$) expresses the time region during which the entanglement can be witnessed by $MU<1$. The witnessed concurrence($C_{EW}$) indicates the entanglement region witnessed by $MU<1$.

\begin{center}
\includegraphics[width=17cm,height=6cm]{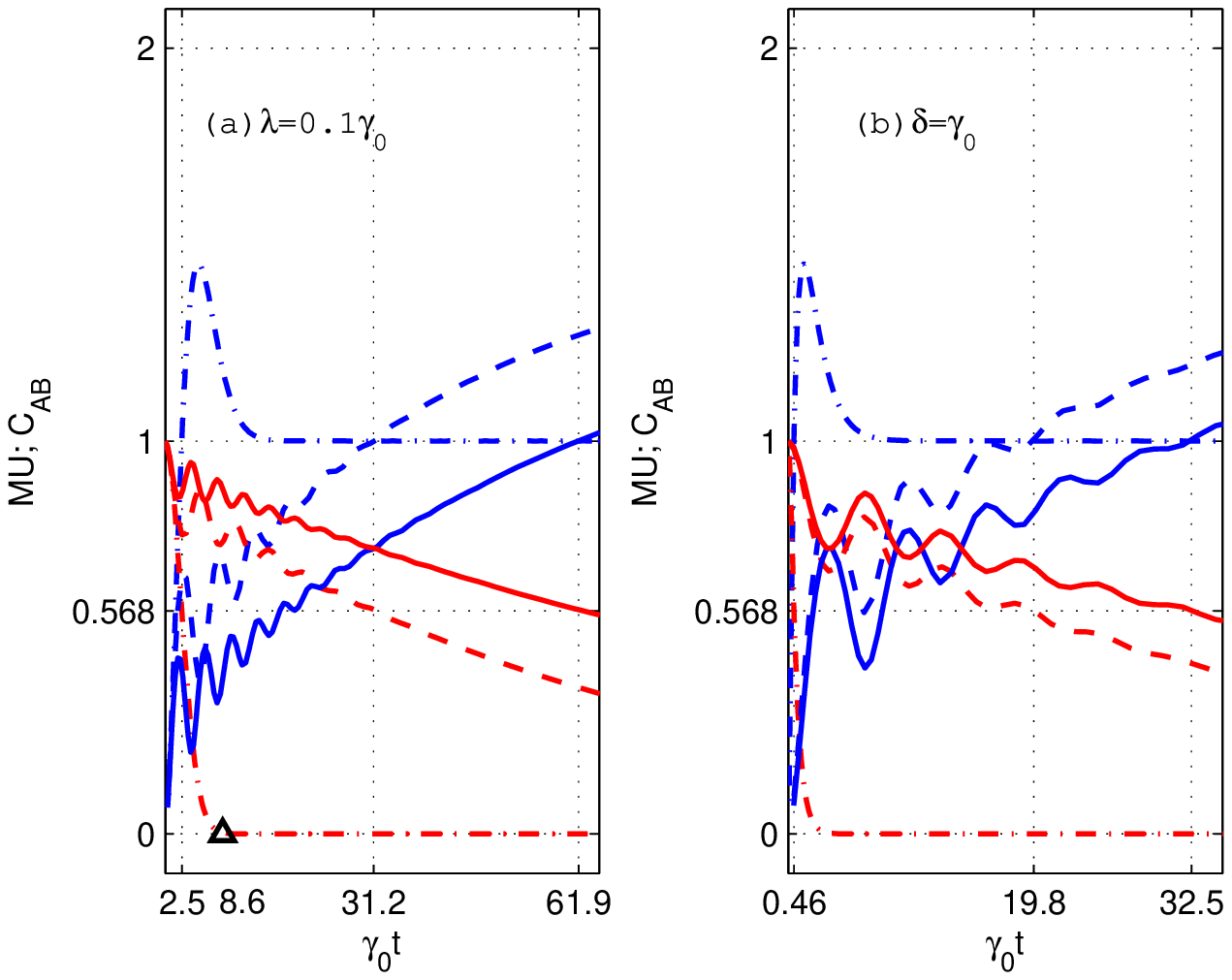}
\parbox{16cm}{\small{\bf Fig1.}
$MU$ and $C_{AB}$ versus $\gamma_{0}t$ in the two identical reservoirs($\lambda_{A}=\lambda_{B}=\lambda$,$\delta_{A}=\delta_{B}=\delta$). $MU$(blue line), $C_{AB}$(red line). (a)the influence of $\delta$ on the entanglement witness in the non-Markovian regime($\lambda=0.1\gamma_{0}$): $\delta=0$(dotted-dashed line), $\delta=1.2\gamma_{0}$(dashed line), $\delta=1.6\gamma_{0}$(solid line); (b)the influence of $\lambda$ on the entanglement witness with detuning($\delta=\gamma_{0}$): $\lambda=5\gamma_{0}$(dotted-dashed line), $\lambda=0.1\gamma_{0}$(dashed line), $\lambda=0.08\gamma_{0}$(solid line).}
\end{center}

In Fig.1, we display the influence of detuning and non-Markovian effect on $MU$ and $C_{AB}$ versus $\gamma_{0}t$ in the two identical reservoirs($\lambda_{A}=\lambda_{B}=\lambda$,$\delta_{A}=\delta_{B}=\delta$). From Fig.1(a), we can see that, in the non-Markovian regime($\lambda=0.1\gamma_{0}$), if $\delta=0$, $C_{AB}$(red dotted-dashed line) quickly decays to 0 and $MU$(blue dotted-dashed line) becomes larger than 1 in a very short time. That is, when $0\leq\gamma_{0}t<2.5$, $MU<1$ and $1\geq C_{AB}>0.568$, the entanglement between $A$ and $B$ can be witnessed by $MU$, while $2.5\leq\gamma_{0}t\leq8.6$, $MU\geq1$ and $0.568\geq C_{AB}\geq0$, thus the entanglement between $A$ and $B$ cannot be witnessed by $MU$ though $A$ is still entangled with $B$. In this case, $C_{EW}$ is from 0.568 to 1.0 and $T_{EW}$ is from 0 to 2.5. A short $T_{EW}$ will restrict the application of entanglement witness in quantum information. When $\delta=1.2\gamma_{0}$, $C_{AB}$(red dashed line) oscillates damply and disappears in a long time, but $MU$(blue dashed line) will be larger than 1 after finite timescales, i.e. when $0\leq\gamma_{0}t<31.2$, $MU<1$ and $1\geq C_{AB}>0.568$, the entanglement between $A$ and $B$ can be witnessed by $MU$, while $\gamma_{0}t\geq31.2$, the entanglement between $A$ and $B$ cannot be witnessed due to $MU\geq1$ though $A$ will be still entangled with $B$ for a long time, so that $T_{EW}\in[0,31.2)$, but it is very interesting that there is still $C_{EW}\in(0.568,1]$. When $\delta=1.6\gamma_{0}$, $C_{AB}$(red solid line) is obviously protected and $MU$(blue solid line) slow rises, $T_{EW}\in[0,61.9)$ and $C_{EW}\in(0.568,1]$. Thus $MU$ can be reduced, $T_{EW}$ can be lengthened but $C_{EW}$ can be effectively protected by increasing the detuning in the non-Markovian regime. Fig.1(b) shows that, in the detuning($\delta=\gamma_{0}$), when $\lambda=5\gamma_{0}$(i.e. in the Markovian regime), $C_{AB}$(red dotted-dashed line) fast decays and $MU$(blue dotted-dashed line) rapidly increases, $T_{EW}\in[0,0.46)$ and $C_{EW}\in(0.568,1]$. When $\lambda=0.1\gamma_{0}$(i.e. with the small non-Markovian effect), $C_{AB}$(red dashed line) reduces and $MU$(blue dashed line) increases, $T_{EW}\in[0,19.8)$ and $C_{EW}\in(0.568,1]$. When $\lambda=0.08\gamma_{0}$(i.e. with the strong non-Markovian effect), $C_{AB}$(red solid line) is effectively protected and $MU$(blue solid line) slow rises, $T_{EW}\in[0,32.5)$ and $C_{EW}\in(0.568,1]$. Hence, in the detuning($\delta=\gamma_{0}$), with $\lambda$ reducing, $MU$ will become small and $T_{EW}$ will extend, but $C_{EW}$ is invariant. Therefore, when both non-Markovian effect and detuning are present simultaneously, increasing detuning and non-Markovian effect can reduce the minimum uncertainty($MU$), lengthen the time of entanglement witnessed($T_{EW}$), and effectively protect the witnessed concurrence($C_{EW}$).

If the two reservoirs have the same spectral width but different detunings, $MU$(blue line) and $C_{AB}$(red line) in Fig.2 exhibit different behaviors from Fig.1. From Fig.2(a), it is found that, in the non-Markovian regime($\lambda_{A}=\lambda_{B}=0.1\gamma_{0}$), when $\delta_{A}=0$ and $\delta_{B}\geq0$, $C_{AB}$ quickly reduces to 0 and $MU$ quickly rises to 2, $T_{EW}$ is very short. With $\delta_{B}$ increasing, $T_{EW}$ has only a little change but $C_{EW}$ becomes clearly narrow. For instance, when $\delta_{B}=0$, $T_{EW}\in[0,2.5)$ and $C_{EW}\in(0.568,1]$, but when $\delta_{B}=4\gamma_{0}$, $T_{EW}\in[0,3.8)$ and $C_{EW}\in(0.663,1]$. Fig.2 (b) indicates that, when $\delta_{A}=0$ and $\delta_{B}=2\gamma_{0}$, the value of $\lambda$ affects $MU$ and $C_{AB}$. With $\lambda$ decreasing, $T_{EW}$ has a little change but $C_{EW}$ becomes also narrow. For example, when $\lambda_{A}=\lambda_{B}=5\gamma_{0}$, $T_{EW}\in[0,0.38)$ and $C_{EW}\in(0.568,1]$, but when $\lambda_{A}=\lambda_{B}=0.05\gamma_{0}$, $T_{EW}\in[0,5.3)$ and $C_{EW}\in(0.652,1]$.

\begin{center}
\includegraphics[width=17cm,height=6cm]{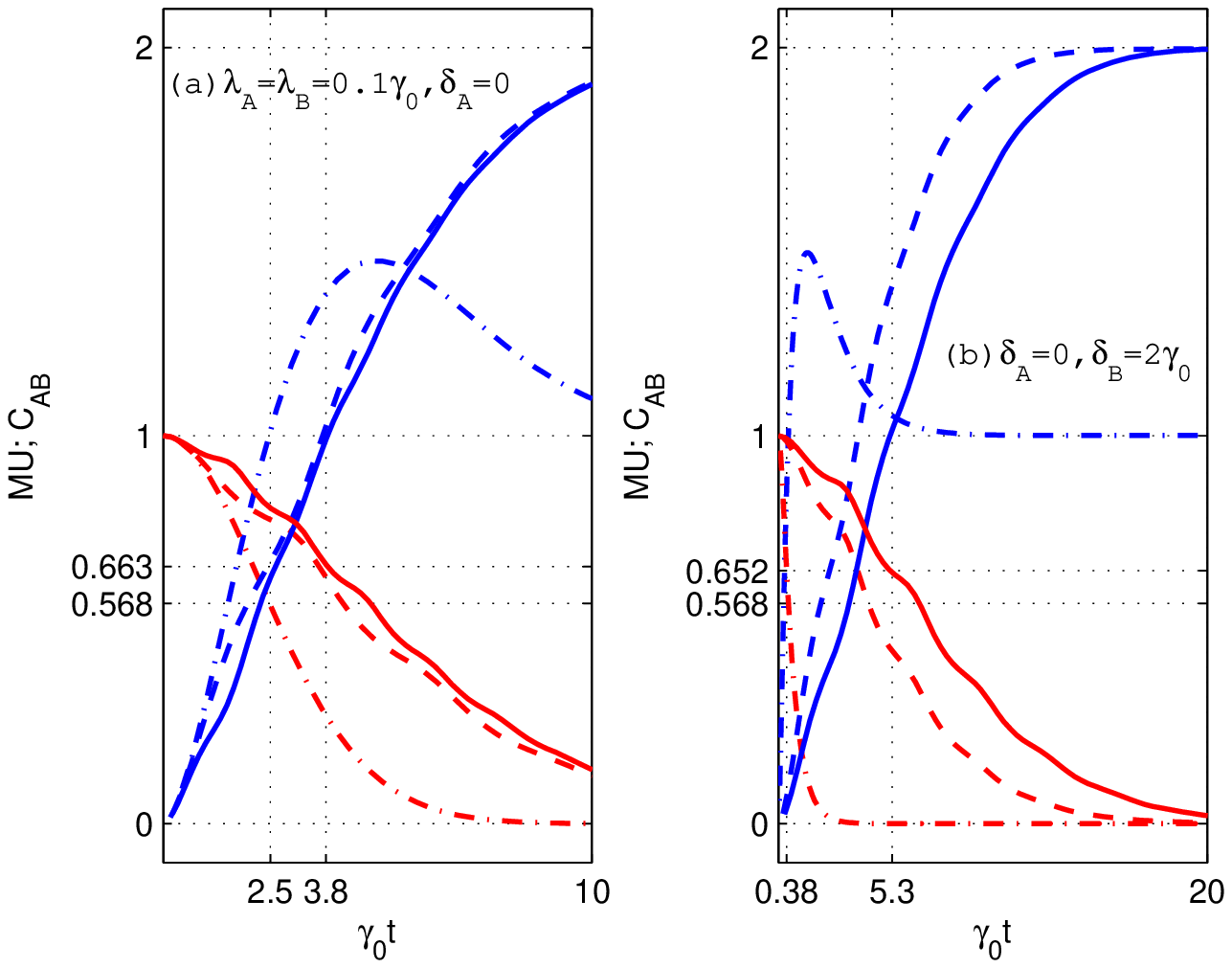}
\parbox{16cm}{\small{\bf Fig2.}
$MU$ and $C_{AB}$ versus $\gamma_{0}t$ in the different reservoirs which have the same spectral width but different detunings. $MU$(blue line), $C_{AB}$(red line). (a)the influence of $\delta_{B}$ on the entanglement witness when $\lambda_{A}=\lambda_{B}=0.1\gamma_{0}$ and $\delta_{A}=0$: $\delta_{B}=0$(dotted-dashed line), $\delta_{B}=2\gamma_{0}$(dashed line), $\delta_{B}=4\gamma_{0}$(solid line); (b)the influence of $\lambda$ on the entanglement witness when $\delta_{A}=0$ and $\delta_{B}=2\gamma_{0}$: $\lambda_{A}=\lambda_{B}=5\gamma_{0}$(dotted-dashed line), $\lambda_{A}=\lambda_{B}=0.1\gamma_{0}$(dashed line), $\lambda_{A}=\lambda_{B}=0.05\gamma_{0}$(solid line).}
\end{center}

Fig.3 depicts $MU$(blue line) and $C_{AB}$(red line) versus $\gamma_{0}t$ when a reservoir is non-Markovian\\
($\lambda_{A}=0.1\gamma_{0}$) and another is Markovian($\lambda_{B}=5\gamma_{0}$). The results show that the detunings $\delta_{A}$ and $\delta_{B}$ hardly affect entanglement witness. $T_{EW}\in[0,0.855)$ and $C_{EW}\in(0.645,1]$ in Fig.3(a), $T_{EW}\in[0,0.838)$ and $C_{EW}\in(0.641,1]$ in Fig.3(b). As a result, $T_{EW}$ and $C_{EW}$ remain unchanged in this case.

\begin{center}
\includegraphics[width=17cm,height=6cm]{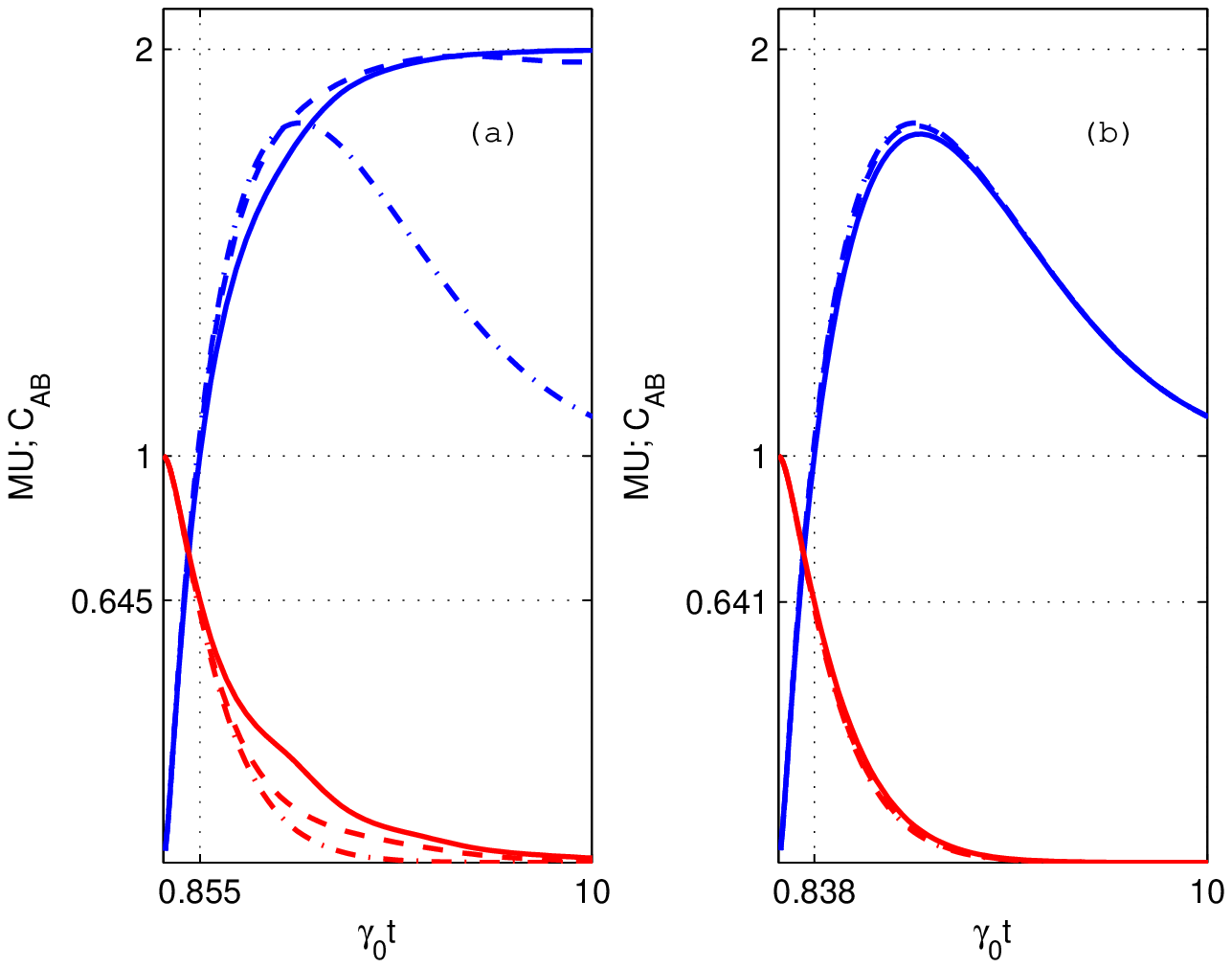}
\parbox{16cm}{\small{\bf Fig3.}
$MU$ and $C_{AB}$ versus $\gamma_{0}t$ in the two different reservoirs that a reservoir is non-Markovian($\lambda_{A}=0.1\gamma_{0}$) and another is Markovian($\lambda_{B}=5\gamma_{0}$). $MU$(blue line), $C_{AB}$(red line). (a)$\delta_{A}=\delta_{B}=0$(dotted-dashed line), $\delta_{A}=\delta_{B}=\gamma_{0}$(dashed line), $\delta_{A}=\delta_{B}=2\gamma_{0}$(solid line); (b)the influence of $\delta_{B}$ on the entanglement witness when $\delta_{A}=0$: $\delta_{B}=0$(dotted-dashed line), $\delta_{B}=\gamma_{0}$(dashed line), $\delta_{B}=2\gamma_{0}$(solid line). Please note that the dotted-dashed line and the dashed line are recovered by the solid line in Fig.3(b).}
\end{center}

Therefore, from the above analysis, we find that, only if the two environments are identically non-Markovian reservoirs, increasing detuning and non-Markovian effect can reduce the minimum uncertainty($MU$), lengthen the time of entanglement witnessed($T_{EW}$), and effectively protect the witnessed concurrence($C_{EW}$).

We can explain the above results using the correlation function $f(t)$ in Eq.~(\ref{EB12}). In Fig.4, we describe $f(t)$ as a function of $\gamma_{0}t$. In the non-Markovian regime($\lambda=0.1\gamma_{0}$), the effect of detuning on $f(t)$ is shown in Fig.4(a). When $\delta=0$, $f(t)$ is always positive and quickly reach a bigger value. The quantum information will very speedily outflow from the atom so that $C_{AB}$ rapidly declines, $MU$ quickly increases and $T_{EW}$ is very short. Nevertheless, when $\delta>0$, $f(t)$ oscillates and its amplitude reduces with $\delta$ increasing. Negative value of $f(t)$ can be understood as the feedback of information from the environment into the atom due to the memory effect of the environment. The amplitude decrease means the atomic decay rate reducing. Only if the two environments are identically non-Markovian reservoirs, the decay of two atoms can slows synchronously and the information is synchronously returned to the two atoms from the environments. In this case, increasing detuning can reduce $MU$, lengthen $T_{EW}$, and effectively protect $C_{EW}$. The influence of the spectral width $\lambda$ on $f(t)$ is shown in Fig.4(b) with detuning($\delta=\gamma_{0}$). When $\lambda=5\gamma_{0}$, $f(t)$ is also always positive and immediately attains a stationary value, which leads to $C_{AB}$ decreasing and $MU$ increasing rapidly. However, when $\lambda<2\gamma_{0}$, $f(t)$ oscillates and its amplitude reduces with $\lambda$ decreasing. At the same time, only if the two non-Markovian reservoirs are identical, $MU$ can be reduced, $T_{EW}$ can be lengthened and $C_{EW}$ can be effectively protected by enhancing the non-Markovian effect.

\begin{center}
\includegraphics[width=17cm,height=6cm]{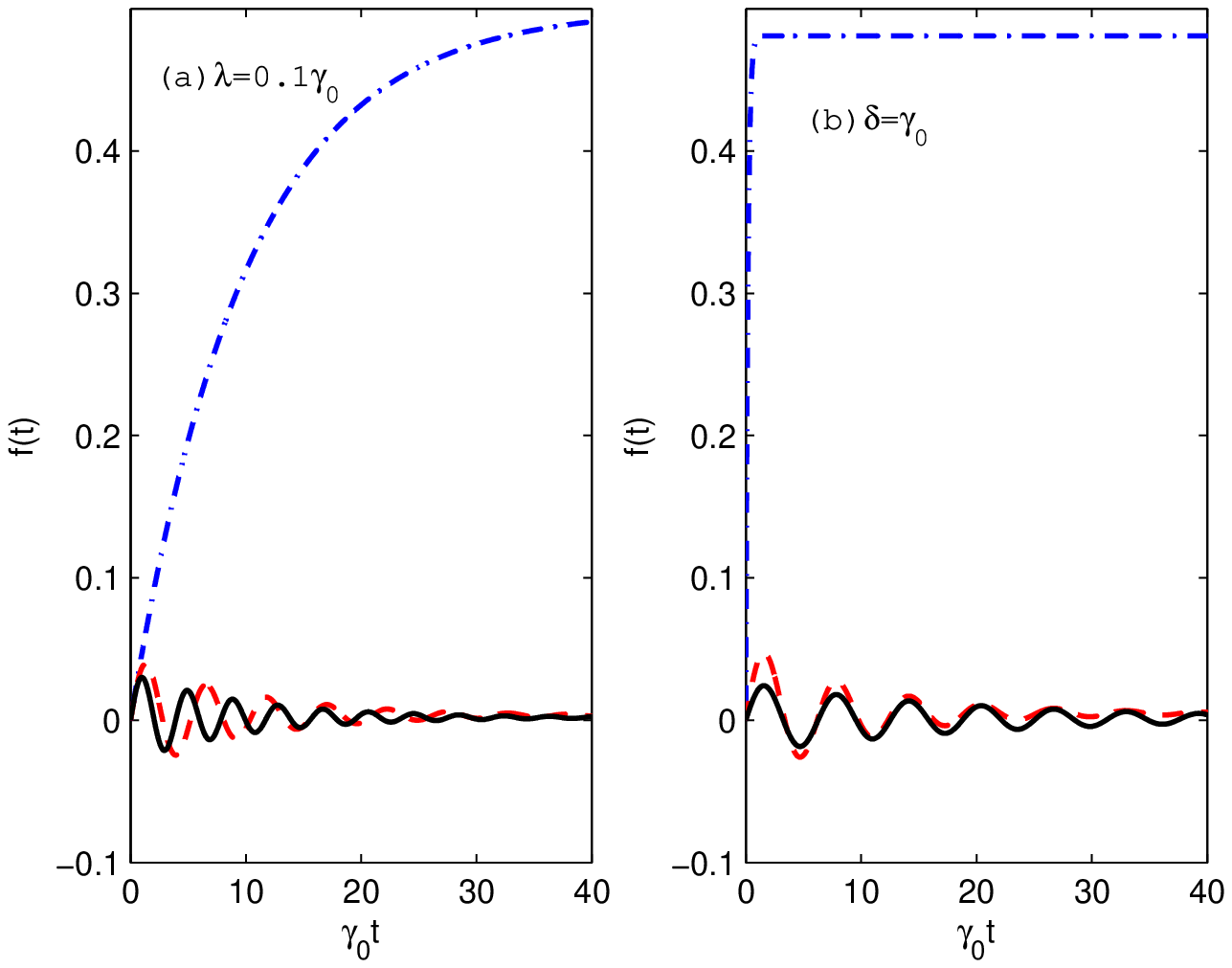}
\parbox{16cm}{\small{\bf Fig4.}
$f(t)$ as a function of $\gamma_{0}t$. (a)the influence of $\delta$ on $f(t)$ in the non-Markovian regime($\lambda=0.1\gamma_{0}$): $\delta=0$(blue dotted-dashed line), $\delta=1.2\gamma_{0}$(red dashed line), $\delta=1.6\gamma_{0}$(black solid line); (b)the influence of $\lambda$ on $f(t)$ with detuning($\delta=\gamma_{0}$): $\lambda=5\gamma_{0}$(blue dotted-dashed line), $\lambda=0.1\gamma_{0}$(red dashed line), $\lambda=0.05\gamma_{0}$(black solid line).}
\end{center}

\section{\fs Conclusion}
In conclusion, we have investigated the quantum entropic uncertainty relation and entanglement witness in the two-atom system coupling with the non-Markovian environments by the time-convolutionless master-equation approach. The influence of non-Markovian effect and detuning on the lower bound of the quantum entropic uncertainty relation and entanglement witness in the presence of quantum memory is discussed in detail. The results show that, only if the two non-Markovian reservoirs are identical, increasing detuning and non-Markovian effect can reduce the lower bound of the entropic uncertainty relation, lengthen the time region during which the entanglement can be witnessed, and effectively protect the entanglement region witnessed by the lower bound of the entropic uncertainty relation. The results can be applied in quantum measurement, entanglement detecting, quantum cryptography task and  quantum information processing.

\section{\fs Acknowledgments}
This work was supported by the Natural Science Foundation of Hunan Province,China(Grant no. 09JJ5001), the Science and Technology Plan of Hunan Province,China(Grant no. 2010FJ3148) and the National Natural Science Foundation of China (Grant No.11374096).

\centerline{\hbox to 8cm{\hrulefill}}

\end{CJK*} 

\begin{thebibliography}{999}
\itemsep=-4pt plus.2pt minus.2pt  
\small
\bibitem{Heisenberg}Heisenberg W 1927 \"{U}ber den anschaulichen Inhalt der quantentheoretischen Kinematik und Mechanik \emph{Z. Phys.} \textbf{43} 172
\bibitem{Bialynicki1}Bialynicki-Birula I 1984 Entropic uncertainty relations \emph{Phys. Lett. A} \textbf{103} 253
\bibitem{Bialynicki2}Bialynicki-Birula I and Madajczyk J L 1985 Entropic uncertainty relations for angular distributions \emph{Phys. Lett. A} \textbf{108} 384
\bibitem{Vaccaro}Vaccaro J A and Bonner R F 1995 Pegg-Barnett phase operators of infinite rank \emph{Phys. Lett. A} \textbf{198} 167
\bibitem{Hirschman}Hirschman I I 1957 A note on entropy \emph{Amer. J. Math.} \textbf{79} 152
\bibitem{Deutsch}Deutsch D 1983 Uncertainty in Quantum Measurements \emph{Phys. Rev. Lett.} \textbf{50} 631
\bibitem{Kraus}Kraus K 1987 Complementary observables and uncertainty relations \emph{Phys. Rev. D} \textbf{35} 3070
\bibitem{Maassen}Maassen H and Uffink J B 1988 Generalized entropic uncertainty relations \emph{Phys. Rev. Lett.} \textbf{60} 1103
\bibitem{Sanchez}S\'{a}nchez-Ruiz J 1993 Entropic uncertainty and certainty relations for complementary observables \emph{Phys. Lett. A} \textbf{173} 233
\bibitem{Renes}Renes J M and Boilean J-C 2009 Conjectured Strong Complementary Information Tradeoff \emph{Phys. Rev. Lett.} \textbf{103} 020402
\bibitem{Berta}Berta M, Christandl M, Colbeck R, Renes J M and Benner B 2010 The Uncertainty Principle in the Presence of Quantum Memory \emph{Nat. Phys.} \textbf{6} 659
\bibitem{Prevedel}Prevedel R, Hamel D R, Colbeck R, Fisher K and Resch K J 2011 Experimental investigation of the uncertainty principle in the presence of quantum memory \emph{Nat. Phys.} \textbf{7} 757
\bibitem{LiC}Li C F, Xu J S, Xu X Y, Li K and Guo G C 2011 Experimental investigation of the entanglement-assisted entropic uncertainty principle \emph{Nat. Phys.} \textbf{7} 752
\bibitem{Tomamichel}Tomamichel M, Lim C C W, Gisin N and Renner R 2012 Tight Finite-Key Analysis for Quantum Cryptography \emph{Nature Commun.} \textbf{3} 634
\bibitem{DiVincenzo}DiVincenzo D P, Horodecki M, Leung D W, Smolin J A and Terhal B M 2004 Locking classical correlations in quantum states \emph{Phys. Rev. Lett.} \textbf{92} 067902
\bibitem{Nataf}Nataf P, Dogan M and Hur K L 2012 Heisenberg uncertainty principle as a probe of entanglement entropy: Application to superradiant quantum phase transitions \emph{Phys. Rev. A} \textbf{86} 043807
\bibitem{Hu1}Hu M-L and Fan H 2012 Quantum-memory-assisted entropic uncertainty principle, teleportation, and entanglement witness in structured reservoirs \emph{Phys. Rev. A} \textbf{86} 032338
\bibitem{Breuer1}Breuer H-P and Petruccione F 2002 \textit{The Theory of Open Quantum Systems}(Oxford University Press, Oxford)
\bibitem{ZouHM}Zou H M, Fang M F and Yang B Y 2013 The squeezing dynamics of two independent atoms by detuning in two non-Markovian environments \emph{Chin. Phys. B} \textbf{22} 120303
\bibitem{Xu1}Xu Z-Y, Yang W L and Feng M 2012 Quantum-memory-assisted entropic uncertainty relation under noise \emph{Phys. Rev. A} \textbf{86} 012113
\bibitem{Pati}Pati A K, Wilde M M, Usha Devi A R, Rajagopal A K and Sudha 2012 Quantum discord and classical correlation can tighten the uncertainty principle in the presence of quantum memory \emph{Phys. Rev. A} \textbf{86} 042105
\bibitem{Ferraro}Ferraro E, Scala M, Migliore R and Napoli A 2009 Non-Markovian dissipative dynamics of two coupled qubits in independent reservoirs: Comparison between exact solutions and master-equation approaches \emph{Phys. Rev. A} \textbf{80} 042112
\bibitem{Sinayskiy}Sinayskiy I, Ferraro E, Napoli A, Messina A and Petruccione F 2009 Non-Markovian dynamics of interacting qubit pair coupled to two independent bosonic baths \emph{J. Phys. A: Math. Theor.} \textbf{42} 485301
\bibitem{Breuer2}Breuer H P, Laine E M and Piilo J 2009 Measure for the Degree of Non-Markovian Behavior of Quantum Processes in Open Systems \emph{Phys. Rev. Lett.} \textbf{103} 210401
\bibitem{dalton}Dalton B J, Barnett S M and Garraway B M 2001 Theory of Pseudomodes in Quantum Optical Processes \emph{Phys. Rev. A} \textbf{64} 053813
\bibitem{Bellomo1}Bellomo B, Franco R Lo and Compagno G 2007 Non-Markovian Effects on the Dynamics of Entanglement \emph{Phys. Rev. Lett.} \textbf{99} 160502
\bibitem{Bellomo2}Bellomo B, Franco R Lo and Compagno G 2008 Entanglement dynamics of two independent qubits in environments with and without memory \emph{Phys. Rev. A} \textbf{77} 032342
\bibitem{ChenJJ}Chen J-J, An J-H, Tong Q-J, Luo H-G and Oh C H 2010 Non-Markovian effect on the geometric phase of a dissipative qubit \emph{Phys. Rev. A} \textbf{81} 022120
\bibitem{Nielsen}Nielsen M A and Chuang I L 2000 Quantum Computation and Quantum Information(Cambridge University Press, Cambridge, England).
\bibitem{Cerf}Cerf N J and Adami C 1997 Negative Entropy and Information in Quantum Mechanics \emph{Phys. Rev. Lett.} \textbf{79} 5194
\bibitem{Vollbrecht}Vollbrecht K G H and Wolf M M 2002 Conditional entropies and their relation to entanglement criteria \emph{arXiv:} quant-ph/0202058
\bibitem{Devetak}Devetak I and Winter A 2005 Distillation of secret key and entanglement from quantum states \emph{Proc. R. Soc. London Ser. A} \textbf{461} 207-235
\bibitem{Wootter}Wootter W K 1998 Entanglement of Formation of an Arbitrary State of Two Qubits \emph{Phys. Rev. Lett.} \textbf{80} 2245

\end{thebibliography}
\end{document}